# Design and Implementation of Washing Machine HUD Using FPGAs

Norman Stites and D.G. Perera

ECE Department, University of Colorado Colorado Springs

Colorado Springs, Colorado, USA

## Abstract

In contemporary digital design education, practical field programmable gate array (FPGA) projects are indispensable for bridging theoretical concepts with real-world applications. This project focuses on developing a hardware-based simulation of a domestic washing machine controller using the Xilinx Spartan-3E development board. A critical component of the design is the graphical heads-up display (HUD), which renders real-time information about the machine's operational state and cycle selections via a VGA interface.

# Section 1: Introduction

## 1.1 Current Scenario

In contemporary digital design education, practical field programmable gate array (FPGA) projects are indispensable for bridging theoretical concepts with real-world applications. This project focuses on developing a hardware-based simulation of a domestic washing machine controller using the Xilinx Spartan-3E development board. The system integrates multiple user input methods—mechanical buttons and a rotary encoder—to emulate real-world interactions, while precise time-based process control ensures accurate management of washing cycles. To enhance user engagement and transparency, a video graphics array (VGA) graphical display is employed to provide real-time visual feedback on the washing machine's operational state. By leveraging the Spartan-3E's configurable logic and peripheral interfaces, this project exemplifies how FPGA platforms enable students to prototype complex embedded systems, fostering hands-on learning in digital design and control systems.

## 1.2 Problem Description and Project Summary

The core challenge of this project lies in designing, implementing, and verifying the VGA signal into a monitor and finite state machine (FSM) capable of orchestrating the sequential stages of a washing cycle, including Fill, Wash, Drain, Rinse, Spin, and Hold. Key technical hurdles include managing robust user interactions through accurate input detection, which requires debouncing mechanical buttons and filtering quadrature signals from the rotary encoder. Precise timing is achieved through dedicated modules that generate operational pulses, ensuring cycle stages adhere to predefined durations. The FSM governs state transitions based on elapsed time, user inputs like the start signal, load size and a safety signal such as door open status. A critical component of the design is the graphical heads-up display (HUD), which renders real-time information about the machine's operational state and cycle selections via a VGA interface. These subsystems debouncing circuits, FSM, timing logic, and VGA controller are integrated into a cohesive top-level module, adhering to the Spartan-3E board's constraints outlined in its user constraint file (UCF). This holistic approach ensures the system operates reliably within the FPGA's resource limits while meeting real-time performance requirements.

## 1.3 Motivation

This project is driven by the goal of deepening understanding into the practical expertise of digital systems design. By translating the operational logic of a household appliance into FPGA hardware, I was able to gain firsthand experience with real-time system design, synchronous control methodologies, and peripheral interfacing. The project challenges students to convert algorithmic descriptions of washing cycles into synthesizable Verilog code, manage parallel processes such as input handling and display refreshing, and implement advanced techniques for signal integrity, debouncing noisy mechanical inputs. Additionally, mastering

VGA signal generation, which is a task requiring strict adherence to VESA video timing specifications prepared me for industry roles involving video interfacing and embedded hardware integration. The Spartan-3E platform, with its cost-effectiveness and academic prevalence, serves as an ideal vehicle for exploring these concepts, equipping learners with skills directly applicable to IoT and industrial automation domains.

## 1.4 Main Goal

The primary objective is to deliver a fully operational FPGA-based washing machine controller that reliably sequences through washing cycle stages, governed by the user inputs and safety conditions. Success hinges on seamless integration of subsystems: the FSM must transition states based on a load size to control timer triggers and door status, the rotary encoder must adjust load sizes without signal jitter, and the VGA HUD must update dynamically to reflect the machine's state. Rigorous evaluation includes functional verification through simulation waveforms, state transitions, timer accuracy, practical validation via hardware demonstrations on the Spartan-3E board, and strict compliance with UCF constraints for input/output mapping and timing. Achieving these milestones will demonstrate the viability of FPGA-based control systems in consumer appliances and reinforce foundational skills in digital design, preparing myself for more advanced projects in embedded systems and hardware design.

# Section 2: Detailed Project Description

## 2.1 System-Level Design

The FPGA-based washing machine controller is a hierarchical digital system designed to emulate the operation of a domestic appliance through synchronized control logic, user input management, and real-time visual feedback. At its core, the system integrates multiple subsystems that collectively automate washing cycles while adhering to timing constraints and user interactions. The Spartan-3E FPGA serves as the central processing unit, coordinating inputs from mechanical buttons and a rotary encoder, managing state transitions via a finite state machine, generating load-dependent timing signals, and driving a VGA display to visualize operational states.

The architecture emphasizes modularity for scalability and maintainability. User inputs, including mechanical buttons for start/reset and door commands and a rotary encoder for load size selection, are processed through an input handler incorporating debouncing circuits and rotary encoder filtering. These inputs are reliably converted into digital signals for FSM management, controlling the sequential progression through washing stages Fill, Wash, Drain, Rinse, Spin, and Hold. A dedicated timing module generates precise, load-dependent durations for each load size, utilizing 32-bit counters to handle small, medium, and large loads with respective timings to the 50MHz configuration clock (CCLK). Concurrently, a VGA heads-up display (HUD) subsystem provides real-time graphical feedback by mapping the FSM's current state onto the VGA output, synchronized accurately by a VGA controller handling horizontal and vertical sync signals. Below we can see the UFC allows us to use only a 3-bit color encoder to output 8 different colors.

```
# VGA Display
NET      "RD"         LOC = "H14" | IOSTANDARD = LVTTL | DRIVE = 8 | SLEW = FAST;
NET      "GR"         LOC = "H15" | IOSTANDARD = LVTTL | DRIVE = 8 | SLEW = FAST;
NET      "BL"         LOC = "G15" | IOSTANDARD = LVTTL | DRIVE = 8 | SLEW = FAST;
NET      "HS"         LOC = "F15" | IOSTANDARD = LVTTL | DRIVE = 8 | SLEW = FAST;
NET      "VS"         LOC = "F14" | IOSTANDARD = LVTTL | DRIVE = 8 | SLEW = FAST;
```

Figure 1. VGA UCF

Table 6-1: 3-Bit Display Color Codes

| VGA_RED | VGA_GREEN | VGA_BLUE | Resulting Color |
|---|---|---|---|
| 0 | 0 | 0 | Black |
| 0 | 0 | 1 | Blue |
| 0 | 1 | 0 | Green |
| 0 | 1 | 1 | Cyan |
| 1 | 0 | 0 | Red |
| 1 | 0 | 1 | Magenta |
| 1 | 1 | 0 | Yellow |
| 1 | 1 | 1 | White |

Figure 2. Spartan 3E Color Decoder

The integrated top-level module (`wm_top.v`) orchestrates all subsystems, ensuring cohesive functionality under the specific constraints defined by the Spartan-3E development board in the user constraint file (`spartan3e.ucf`).

## 2.2 Design Flow

The project adheres to a structured, iterative design process, beginning with initial requirements analysis that clearly defines washing cycle stages, timing parameters, and user interface specifications, while considering FPGA constraints such as the 50 MHz clock and limited logic slices. Submodules like debouncing circuits, FSM logic, timing logic, and graphical

HUD management were methodically developed in Verilog HDL, emphasizing synthesizable code and synchronous design practices.

Following individual module development, extensive simulation and verification through testbenches ensured accurate functionality such as confirming correct FSM transitions upon receiving specific input signals. Timing analysis verified that critical paths adhered strictly to the 50 MHz clock constraints for most of the modules, and a 25MHz clock divider was used for the VGA signals per the VESA specifications for 640x480@60 Hz calls for a pixel clock of 25.175 MHz even though we don't exactly have this frequency the VGA input can down clock and latch onto a slow clock if needed. The Xilinx ISE toolchain facilitated synthesis, mapping, and bitstream generation, guided by the custom designed UCF file assigning physical pins for buttons, rotary encoder signals, buzzer, and VGA outputs.

Post-synthesis, the system underwent comprehensive hardware validation, ensuring rotary encoder responsiveness, VGA display accuracy, and robust functionality of the safety feature like pausing the spin cycle when the door opens. Challenges such as managing concurrent processes, signal integrity for mechanical inputs, and resource optimization were systematically addressed. Debouncing circuits using shift registers effectively mitigated mechanical noise, careful FSM encoding reduced complexity, and precise synchronization in the VGA subsystem-maintained display integrity.

Overall, this systematic and iterative design flow from conceptualization through simulation, synthesis, and hardware validation demonstrates the effective use of FPGA platforms in bridging theoretical concepts with practical embedded systems. This robust framework paves the way for future enhancements, including advanced fault detection capabilities or IoT integration.

# Section 3: Detailed Hardware Design

## 3.1 System Block Diagram

The hardware architecture revolves around the Spartan-3E FPGA, which integrates multiple subsystems to manage washing machine operations effectively. The Spartan-3E FPGA serves as the central processing unit, receiving inputs from mechanical buttons and a rotary encoder, managing state transitions through a finite state machine (FSM), generating load-dependent timing signals, and controlling a VGA display for visualizing operational states along with an attached buzzer for audio completion feedback. The FPGA is further responsible for outputting control signals to actuators such as the motor and pump, enabling physical operation if it was connected to, and could drive, those devices.

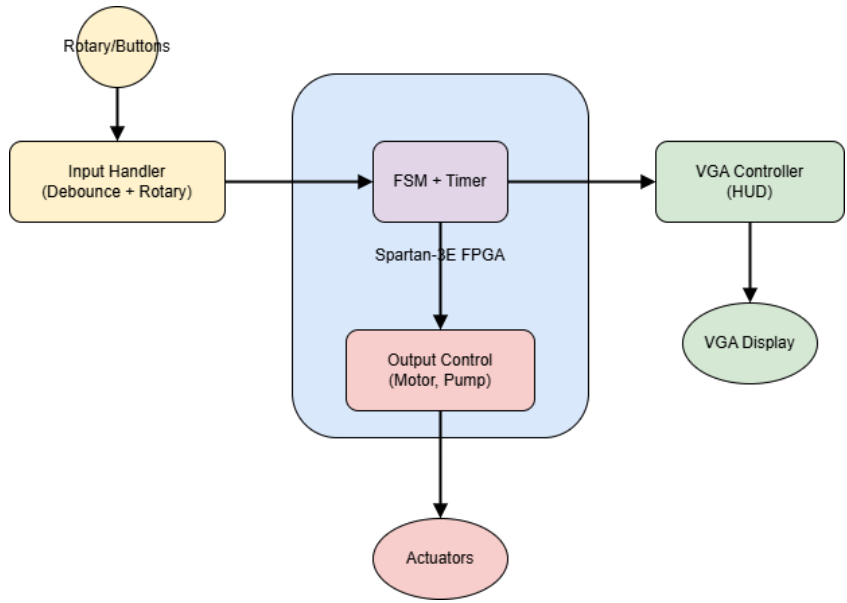

## 3.2 Hardware Components and Peripherals

The central hardware component, the Spartan-3E FPGA board, utilizes a 50 MHz onboard clock that drives all state logic. GPIO pins interface with external peripherals, including VGA output signals (RGB, HSYNC, VSYNC), rotary encoder inputs (ROTA, ROTB), and button inputs (Start, Reset, Door). Light emitting diodes (LEDs) are utilized to indicate the operation of the different actuator devices.

A VGA compatible display with 640×480 resolution is driven by a 25 MHz clock derived from the FPGA's primary clock using a dedicated clock divider. A resistor DAC converts the digital 3-bit red, green, and blue (RGB) signals into analog voltages suitable for VGA output. Mechanical inputs include a rotary encoder providing quadrature signals decoded into clockwise or counterclockwise rotation commands for load size selection, coupled with debouncing circuits employing shift registers to filter out mechanical noise.

## 3.3 Design Rationale

Generating precise VGA signals required careful alignment of horizontal synchronization (HSYNC) and vertical synchronization (VSYNC) pulses. A dedicated VGA synchronization module (`vga_sync.v`) generates these signals by accurately counting horizontal and vertical pixel positions. This module synchronizes pixel data for the graphical heads-up display (HUD), which dynamically maps the current FSM states to color-coded screen regions, clearly indicating operational stages to the user.

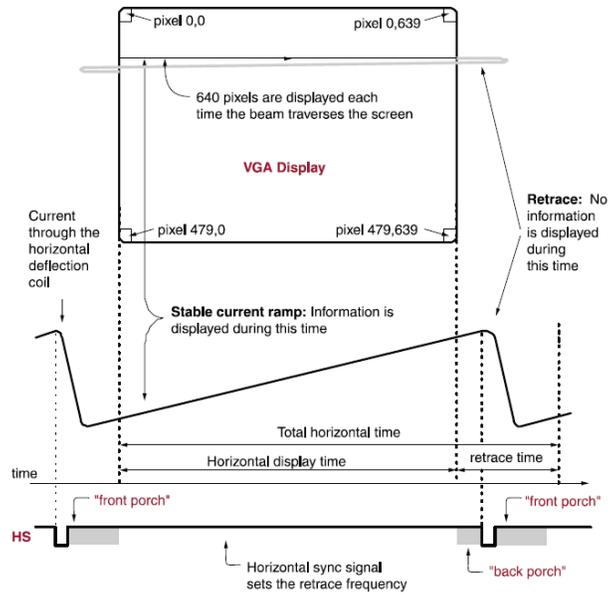

Figure 3. CRT Timing Example

Handling noisy quadrature signals from the rotary encoder posed challenges, prompting the development of a specialized rotary filter module (`rotary_filter.v`). This module decodes quadrature signals into clean pulses, reliably detecting direction based on the relative phase of the rotary signals, ensuring accurate load size selection.

Transitioning FSM logic from simulation to hardware revealed mismatches due to unaccounted timing delays. These were resolved by scaling timer counters to real-world timing constraints and implementing synchronizers to safely bridge signals between the FSM's 50 MHz domain and the VGA's 25 MHz domain. Debugging processes included visual state feedback via LEDs to quickly diagnose and debug stuck or improperly transitioned states.

## 3.4 Key Problems and Solutions

Initial VGA implementation faced synchronization issues due to misaligned HSYNC and VSYNC pulses, resolved by adjusting counter thresholds within the `vga_sync.v` module to match the standard VGA timing specifications.

| Symbol | Parameter | Vertical Sync | | | Horizontal Sync | |
|---|---|---|---|---|---|---|
| | | Time | Clocks | Lines | Time | Clocks |
| $T_S$ | Sync pulse time | 16.7 ms | 416,800 | 521 | 32 μs | 800 |
| $T_{DISP}$ | Display time | 15.36 ms | 384,000 | 480 | 25.6 μs | 640 |
| $T_{PW}$ | Pulse width | 64 μs | 1,600 | 2 | 3.84 μs | 96 |
| $T_{FP}$ | Front porch | 320 μs | 8,000 | 10 | 640 ns | 16 |
| $T_{BP}$ | Back porch | 928 μs | 23,200 | 29 | 1.92 μs | 48 |

Figure 4. 640x480 Mode VGA Timings

Rotary encoder jitter initially caused multiple increments per physical detent, which was effectively mitigated by introducing an input synchronizer based on the default Xilinx Spartan 3E demo code.

```
reg [1:0] rotary_sync;
reg rotary_q1, rotary_q2;

always @(posedge clk) rotary_sync <= {rotary_b, rotary_a};

	always @(posedge clk) begin
		case (rotary_sync)
			2'b00: {rotary_q1, rotary_q2} <= {1'b0, rotary_q2};
			2'b01: {rotary_q1, rotary_q2} <= {rotary_q1, 1'b0};
			2'b10: {rotary_q1, rotary_q2} <= {rotary_q1, 1'b1};
			2'b11: {rotary_q1, rotary_q2} <= {1'b1, rotary_q2};
		endcase
	end

reg delay_rotary_q1;
always @(posedge clk) delay_rotary_q1 <= rotary_q1;
assign rotary_event = (rotary_q1 && !delay_rotary_q1); assign direction = rotary_q2;
```

This causes both asynchronous encoder inputs (rotary_a, rotary_b) to be sampled on the system clock to avoid metastability. This registers all downstream logic to see a stable, glitch-free encoder state as well as detect the direction of the turn, counterclockwise or clockwise.

The rotary_filter module works by first registering the two asynchronous encoder signals with the clock. On each rising edge of the clock it samples ROTA and ROTB into a two-bit register called rotary_sync. Inside a second clocked process it watches rotary_sync as a simple quadrature decoder: whenever rotary_sync is "00" it clears rotary_q1 (arming for the next detent), and whenever it becomes "11" it sets rotary_q1 (marking arrival at a detent). Simultaneously, when rotary_sync is "01" it clears rotary_q2 and when it is "10" it sets rotary_q2, thereby remembering which of the two intermediate quadrature states was traversed—and thus the direction of rotation. A delayed copy of rotary_q1 is held in delay_rotary_q1, and by comparing the current and delayed versions (rotary_q1 && !delay_rotary_q1) the module emits a single-clock-wide pulse on rotary_event each time a new detent is reached. The direction output simply reflects the value of rotary_q2, so every clean pulse is accompanied by a static bit indicating clockwise or counterclockwise motion.

In the first version of my FSM I treated the DOOR push-button exactly like any other state-driving signal and registered it inside my next-state logic. But when I would press the button the DOOR bit would latch high and stay high until a reset or reprogram would clear it. That meant as soon as the machine entered the HOLD state it never saw DOOR go low again, so it simply sat there forever. By moving to an edge-detect approach synchronizing the raw button input into the clock domain and just had to hold the button to simulate the door being open. The FSM only sees the high on DOOR when the button is actuated. That pulse drives the transition into HOLD, but because it's only asserted until released, the FSM can then progress normally out of HOLD back into the SPIN cycle.

Ultimately I ended up scrapping the idea of a push button door and implemented it on the switch, but in order to do this I added a always block on the top layer that only allowed it to be "opened" during the spin cycle.

```
always @(posedge clk_25MHz or posedge BTNS) begin
    if (BTNS) begin
        SW3_block <= 1'b0;
    end
    else if (washing_state == 4'b0111 || washing_state == 4'b1000) begin
        SW3_block <= SW3_sync;
    end
    else
        SW3_block <= 1'b0;
end
```

The last real problem I had was with attaching the buzzer, I saved this for last, but was taken aback by the UCF addition. I originally thought that I could just attach the IO pin to the UCF but that didn't work. I forgot that I need to add in the additional parameters that go with the IO pin.

NET "BUZZ" LOC = "D7" | IOSTANDARD = LVTTL | SLEW = SLOW | DRIVE = 8;

Once I correctly added in the IOSTANDARD to 3.3v, SLEW, and DRIVE to 8 mA, I was able to finally use the buzzer.

## 3.5 Integration and Validation

The top-level module (`wm_top.v`) integrates all subsystems, harmonizing debounced mechanical inputs and rotary signals with the FSM. It orchestrates control signals for actuators and outputs coherent VGA display signals. Integration tests verified correct operation through static and dynamic VGA outputs, validating the display's accuracy. The rotary encoder's functionality was confirmed by observing incremental and decremental load size indications via LEDs. Additionally, FSM behavior was rigorously tested, verifying correct state transitions under various input conditions, including simulated door-open scenarios.

# Section 4: Discussion, Conclusion, and Future Work

## 4.1 Discussion

The FPGA-based washing machine controller successfully demonstrates the integration of digital logic design, real-time input handling, and graphical feedback on the Spartan-3E platform. The finite state machine (FSM) effectively transitions through washing stages, as validated by rigorous simulation and practical hardware tests. The user-centric approach, featuring a rotary encoder for intuitive load selection and a VGA-based graphical heads-up display (HUD), significantly enhances user interaction by bridging the hardware's functional capabilities with user-friendly feedback.

Resource management was particularly noteworthy, as the design efficiently optimized logic utilization, ensuring effective use of the Spartan-3E's limited resources (1,728 logic slices). Technical insights gained include effective signal debouncing and integrity management, where shift-register debouncing reliably eliminated mechanical input inconsistencies, critical for stable operation. Additionally, achieving precise VGA synchronization, specifically adhering to strict horizontal (31.77 µs) and vertical (16.68 ms) sync intervals, ensured stable and accurate graphical outputs. Furthermore, discrepancies between simulation environments and real hardware timing highlighted the necessity and importance of conducting post-synthesis timing analysis to accurately reflect physical propagation delays and system behavior.

Furthermore, I had quite a bit of warnings, but most of them were due to a bit of lazy coding, in that I didn't loop my counters when they got to whatever number they needed to go to. This causes the IDE to see them being able to overflow, which produced the warning. But it was not an issue with the functionality of the design. And the IDE would automatically truncate them to avoid overflow. In a polished product this would have been corrected.

## 4.2 Conclusion

This project highlights the capabilities of the Spartan-3E FPGA platform for prototyping complex embedded systems, effectively integrating control logic, user interfaces, and essential safety mechanisms. By successfully implementing theoretical concepts, including finite state machines, clock domain management, and peripheral interfacing, the design not only reinforces educational objectives but also aligns closely with industrial practices—particularly in features such as automatically pausing operation upon door opening events.

Functional verification using testbenches validated FSM state transitions and timer accuracy, while real-world hardware validation demonstrated the responsiveness and accuracy of the system through visual indicators like LEDs and the VGA output. Efficient resource utilization, evidenced by occupying only 65% of logic slices and 10% of block RAMs, confirmed adherence to effective design practices.

## 4.3 Future Work

To further enhance functionality, usability, and scalability, several enhancements and additions are proposed. Advanced HUD features, such as text rendering via font ROMs and dynamic progress bars animated by pixel counters, can significantly improve the visual interface. Integration with touchscreen displays would further advance user interactivity, potentially replacing the current VGA interface.

Incorporating IoT connectivity, such as integrating Wi-Fi modules like the ESP8266, would allow remote monitoring and control capabilities, significantly expanding the system's functionality and accessibility. Additionally, energy monitoring capabilities could provide valuable insights into water and power usage, promoting efficiency and sustainability.

Enhanced safety and diagnostic features could include incorporating fault detection mechanisms using sensors for water levels and motor temperatures to trigger automatic error states, along with implementing self-test modes to verify actuator functionality upon startup. Hardware upgrades, such as migrating the design to modern FPGAs like the Artix-7, would provide additional resources, including DSP slices and block RAM, enhancing performance and scalability. Employing PWM for motor control instead of direct GPIO signals would facilitate more precise and energy-efficient actuator operations.

Also, energy efficiency can be significantly improved through the implementation of load sensing capabilities using current sensors, thus automating load detection and eliminating manual inputs. Introducing sleep modes would reduce energy consumption during idle periods, aligning the design with contemporary energy efficiency standards.

This work is inspired by the digital design research group at UCCS. This group has done extensive work in FPGA-based architectures, techniques, and associated models. Their analyses [10],[11] shows that FPGA-based systems are currently the best option to support applications and algorithms, such as the ones presented in this report. Also, their previous work on FPGA-based accelerators, architectures, and techniques for various compute and data-intensive applications, including data analytics/mining [12],[13],[14],[15],[16],[17],[18],[19],[20],[21]; control systems [22],[23],[24],[25],[26],[27]; cybersecurity [28],[29],[30]; machine learning [31],[32],[33],[34],[35],[36],[37]; communications [38],[39]; edge computing [40],[41]; bioinformatics [42],[43]; and neuromorphic computing [44],[45]; demonstrated that FPGA-based systems are the best avenue to support and accelerate complex algorithms.

Also as future work, we are planning to investigate hardware optimization techniques, such as parallel processing architectures (similar to [33],[46],[47],[48]), partial and dynamic reconfiguration traits (as stated in [49],[50],[51]) and architectures (similar to [29],[52],[53],[54]), HDL code optimization techniques (as stated in [55],[56]),  and multi-ported memory architectures (similar to [57],[58],[59],[60]), to further enhance the performance metrics of FPGA-based architectures, while considering the associated tradeoffs.

## 4.4 Final Remarks

This project exemplifies the flexibility and power of FPGA technology in embedded system prototyping, providing a robust platform for academic experimentation and industrial innovation. By continuously addressing current limitations and implementing proposed enhancements, the washing machine controller has the potential to evolve into a sophisticated, commercially viable product. This progression highlights the transformative capabilities of digital logic design in practical, everyday technological applications.